\documentclass[twocolumn,showpacs,preprintnumbers]{revtex4}%
\usepackage{amssymb}
\usepackage{amsmath}
\usepackage{graphicx}
\usepackage{dcolumn}
\usepackage{bm}
\usepackage{amsfonts}%
\begin{document}
\title{Displacement Detection with a Vibrating RF SQUID: Beating the Standard Linear Limit}
\author{Eyal Buks, Stav Zaitsev, Eran Segev, Baleegh Abdo}
\affiliation{Department of Electrical Engineering, Technion, Haifa 32000 Israel}
\author{M. P. Blencowe}
\affiliation{Department of Physics and Astronomy, Dartmouth College, Hanover, New Hampshire
03755, USA}
\date{\today }

\begin{abstract}
We study a novel configuration for displacement detection consisting of a
nanomechanical resonator coupled to both, a radio frequency superconducting
interference device (RF SQUID) and to a superconducting stripline resonator.
We employ an adiabatic approximation and rotating wave approximation and
calculate the displacement sensitivity. We study the performance of such a
displacement detector when the stripline resonator is driven into a region of
nonlinear oscillations. In this region the system exhibits noise squeezing in
the output signal when homodyne detection is employed for readout. We show
that displacement sensitivity of the device in this region may exceed the
upper bound imposed upon the sensitivity when operating in the linear region.
On the other hand, we find that the high displacement sensitivity is
accompanied by a slowing down of the response of the system, resulting in a
limited bandwidth.

\end{abstract}
\maketitle





\section{Introduction}

Resonant detection is a widely employed technique in a variety of
applications. A detector belonging to this class typically consists of a
resonator, which is characterized by a resonance frequency $\omega_{0}$ and
characteristic damping rates. Detection is achieved by coupling the measured
physical parameter of interest, denoted as $x$, to the resonator in such a way
that $\omega_{0}$ becomes effectively $x$ dependent, that is $\omega
_{0}=\omega_{0}\left(  x\right)  $. In such a configuration $x$ can be
measured by externally driving the resonator, and monitoring its response as a
function of time by measuring some output signal $X\left(  t\right)  $. Such a
scheme allows a sensitive measurement of the parameter $x$, provided that the
average value of $X\left(  t\right)  $, which is denoted as $X_{0}$, strongly
depends on $\omega_{0}$, and provided that $\omega_{0}$, in turn, strongly
depends on $x$. These dependencies are characterized by the responsivity
factors $R=\left\vert \partial X_{0}/\partial\omega_{0}\right\vert $ and
$\left\vert \partial\omega_{0}/\partial x\right\vert $ respectively. Resonant
detection has been employed before for mass detection \cite{Ekinci_2682},
quantum state readout of a quantum bit
\cite{Wallraff_162,Lupascu_127003,Johansson_S901,Lee_144505}, detection of
gravitational waves \cite{Barish_44}, and in many other applications.

In general, any detection scheme employed for monitoring the parameter of
interest $x$ can be characterized by two important figures of merit. The first
is the minimum detectable change in $x$, denoted as $\delta x$. This parameter
is determined by the above mentioned responsivity factors, the noise level,
which is usually characterized by the spectral density of $X\left(  t\right)
$, and by the averaging time $\tau$ employed for measuring the output signal
$X\left(  t\right)  $. The second figure of merit is the ring-down time
$t_{\mathrm{RD}}$, which is a measure of the detector's response time to a
sudden change in $x$.

In general, the minimum detectable change $\delta x$ is proportional to the
square root of the available bandwidth, namely, to $\left(  2\pi/\tau\right)
^{1/2}$. It is thus convenient to characterize the sensitivity by the minimum
detectable change $\delta x$ per square root of bandwidth, which is given by
$P_{x}=\delta x/\left(  2\pi/\tau\right)  ^{1/2}$. Under some conditions,
which will be discussed below in detail, the smallest possible value of
$P_{x}$ is given by \cite{Ekinci_2682,Cleland_235}%
\begin{equation}
P_{x}^{\mathrm{SLL}}=\left\vert \frac{\partial\omega_{0}}{\partial
x}\right\vert ^{-1}\left(  \frac{\omega_{0}}{2Q}\frac{k_{B}T}{U_{0}}%
\frac{\hbar\omega_{0}}{2k_{B}T}\coth\frac{\hbar\omega_{0}}{2k_{B}T}\right)
^{1/2}\ , \label{P_x linear}%
\end{equation}
where $k_{B}T$ is the thermal energy, $U_{0}$ is the energy stored in the
resonator, and $Q$ is the quality factor of the resonator. One of the
assumptions, which are made in order to derive Eq. (\ref{P_x linear}), is that
the response of the resonator is linear. We therefore refer to the value of
$P_{x}$ given by Eq. (\ref{P_x linear}) as the \textit{standard linear limit}
(SLL) of resonant detection. Under the same conditions and assumptions, the
ring-down time is given by%
\begin{equation}
t_{\mathrm{RD}}=\frac{Q}{\omega_{0}}\ . \label{t_RD linear}%
\end{equation}

As can be seen from Eq. (\ref{P_x linear}), sensitivity enhancement can be
achieved by increasing $Q$, however, this unavoidably will be accompanied by
an undesirable increase in the ring-down time (see Eq. (\ref{t_RD linear})),
namely, slowing down the response of the system to changes in $x$. Moreover,
Eq. (\ref{P_x linear}) apparently suggests that unlimited reduction in $P_{x}$
can be achieved by increasing $U_{0}$ by means of increasing the drive
amplitude. Note however that Eq. (\ref{P_x linear}), which was derived by
assuming the case of linear response, is not applicable in the nonlinear
region. Thus, in order to characterize the performance of the system when
nonlinear oscillations are excited by an intense drive, one has to generalize
the analysis by taking nonlinearity into account
\cite{Lupascu_127003,Wang_025008,Santamore_052105,Cleland_235,Dykman_265,Buke_023815}%
.

In the present paper we theoretically study a novel configuration for resonant
detection of displacement of a nanomechanical resonator
\cite{Knobel_291,LaHaye_74}, which is coupled to both, a radio frequency
superconducting interference device (RF SQUID) and to a superconducting
stripline resonator. Similar configurations have been studied recently in
\cite{Zhou_0605017,Xue_35,Buks_174504,Buks_0610158,Blencowe_0704_0457,Wang_0704_2462}%
. We employ an adiabatic approximation and a rotating wave approximation (RWA)
to simplify the analysis and calculate the displacement sensitivity. We first
consider the case where the response of the stripline resonator is linear and
reproduce Eqs. (\ref{P_x linear}) and (\ref{t_RD linear}) in this limit.
However, we find that the response becomes nonlinear at a relatively low input
drive power. Next, we show that $P_{x}$ in the nonlinear region may become
significantly smaller than the value given by Eq. (\ref{P_x linear}),
exceeding thus the SLL imposed upon the sensitivity when operating in the
linear region. On the other hand, we find that the enhanced displacement
sensitivity is accompanied by a slowing down of the response of the system,
resulting in a limited bandwidth, namely, a ring-down time much longer than
the value given by Eq. (\ref{t_RD linear}).

\section{The Device}

\begin{figure}
[ptb]
\begin{center}
\includegraphics[
height=1.7729in,
width=3.2396in
]%
{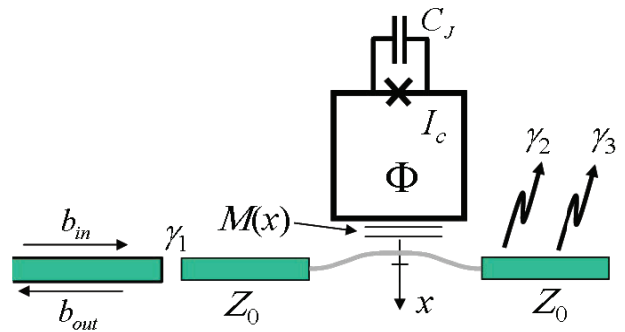}%
\caption{(Color online) The device consists of a nanomechanical resonator
coupled to both, a radio frequency superconducting interference device (RF
SQUID) and to a superconducting stripline resonator.}%
\label{device}%
\end{center}
\end{figure}

The device, which is schematically shown in Fig. \ref{device}, consists of an
RF SQUID inductively coupled to a stripline resonator. The loop of the RF
SQUID, which has a self inductance $\Lambda$, is interrupted by a Josephson
junction (JJ) having a critical current $I_{c}$ and a capacitance $C_{J}$. A
perpendicularly applied magnetic field produces a flux $\Phi_{e}$ threading
the loop of the RF SQUID. The stripline resonator is made of two identical
stripline sections of length $l/2$ each having inductance $L_{T}$ and
capacitance $C_{T}$ per unit length and characteristic impedance $Z_{T}%
=\sqrt{L_{T}/C_{T}}$. The stripline sections are connected by a doubly clamped
beam made of a narrow strip, which is freely suspended and allowed to
oscillate mechanically. We assume the case where the fundamental mechanical
mode vibrates in the plane of the figure and denote the amplitude of this
flexural mode as $x$. Let $m$ be the effective mass of the fundamental
mechanical mode, and $\omega_{m}$ its angular resonance frequency. The
suspended mechanical beam is assumed to have an inductance $L_{b}$
(independent of $x$) and a negligible capacitance to ground. The coupling
between the mechanical resonator and the rest of the system originates from
the dependence of the mutual inductance $M$ between the inductor $L_{b}$ and
the RF SQUID loop on the mechanical displacement $x$, that is $M=M\left(
x\right)  $.%

\subsection{Transmission Line Resonator}

The transmission line is assumed to extend from $y=-l/2$ to $y=l/2$, and the
lumped inductance $L_{b}$ is located at $y=0$. Consider the case where only
the fundamental mode of the resonator is driven. Disregarding all other modes
we express the voltage $\mathcal{V}\left(  y,t\right)  $ and current
$\mathcal{I}\left(  y,t\right)  $ along the transmission line as%
\begin{align}
\mathcal{V}\left(  y,t\right)   &  =\left\{
\begin{array}
[c]{cc}%
\dot{\varphi}\frac{L_{T}\cos\left[  \kappa\left(  y+\frac{l}{2}\right)
\right]  }{\kappa L_{b}\sin\frac{\kappa l}{2}} & y<0\\
-\dot{\varphi}\frac{L_{T}\cos\left[  \kappa\left(  y-\frac{l}{2}\right)
\right]  }{\kappa L_{b}\sin\frac{\kappa l}{2}} & y>0
\end{array}
\right.  \;,\label{V fm}\\
\mathcal{I}\left(  y,t\right)   &  =\left\{
\begin{array}
[c]{cc}%
\varphi\frac{\sin\left[  \kappa\left(  y+\frac{l}{2}\right)  \right]  }%
{L_{b}\sin\frac{\kappa l}{2}} & y<0\\
-\varphi\frac{\sin\left[  \kappa\left(  y-\frac{l}{2}\right)  \right]  }%
{L_{b}\sin\frac{\kappa l}{2}} & y>0
\end{array}
\right.  \;, \label{I fm}%
\end{align}
where $\varphi$ represents the flux in the lumped inductor at $y=0$. The value
of $\kappa$ is determined by applying Faraday's law to the lumped inductor at
$y=0$ \cite{Blencowe_0704_0457}%
\begin{equation}
\cot\frac{\kappa l}{2}=-\frac{\kappa l}{2}\frac{L_{b}}{L_{T}l}\;.
\label{cot(kappa*l/2)}%
\end{equation}
For the fundamental mode the solution is in the range $\pi\leq\kappa l\leq
2\pi$ (see Fig. \ref{vartheta}).

\subsection{Inductive Coupling}

The total magnetic flux $\Phi$ threading the loop of the RF SQUID is given by%
\begin{equation}
\Phi=\Phi_{e}+\Phi_{i}\ , \label{Phi}%
\end{equation}
where $\Phi_{i}$ represents the flux generated by both, the circulating
current in the RF SQUID $I_{s}$ and by the current in the suspended mechanical
beam $I_{b}$%
\begin{equation}
\Phi_{i}=I_{s}\Lambda+MI_{b}\ , \label{Phi_i}%
\end{equation}
where $\Lambda$ is the self inductance of the loop. Similarly, the magnetic
flux $\varphi$ in the inductor $L_{b}$ is given by%
\begin{equation}
\varphi=I_{b}L_{b}+MI_{s}\ . \label{varphi}%
\end{equation}
Inverting these relations yields%
\begin{align}
I_{s}  &  =\frac{L_{b}\Phi_{i}-M\varphi}{\Lambda L_{b}\left(  1-\alpha_{M}%
^{2}\right)  }\ ,\label{I_s}\\
I_{b}  &  =\frac{\Lambda\varphi-M\Phi_{i}}{\Lambda L_{b}\left(  1-\alpha
_{M}^{2}\right)  }\ , \label{I_b}%
\end{align}
where%
\begin{equation}
\alpha_{M}=\frac{M}{\sqrt{\Lambda L_{b}}}\;. \label{alpha_M}%
\end{equation}

The gauge invariant phase across the Josephson junction $\theta$ is given by%
\begin{equation}
\theta=2\pi n-\frac{2\pi\Phi}{\Phi_{0}}\ ,
\end{equation}
where $n$ is an integer and $\Phi_{0}=h/2e$ is the flux quantum. We set $n=0$,
since observable quantities do not depend on $n$.

\subsection{Capacitive and Inductive Energies}

\begin{figure}
[ptb]
\begin{center}
\includegraphics[
height=2.5374in,
width=3.2396in
]%
{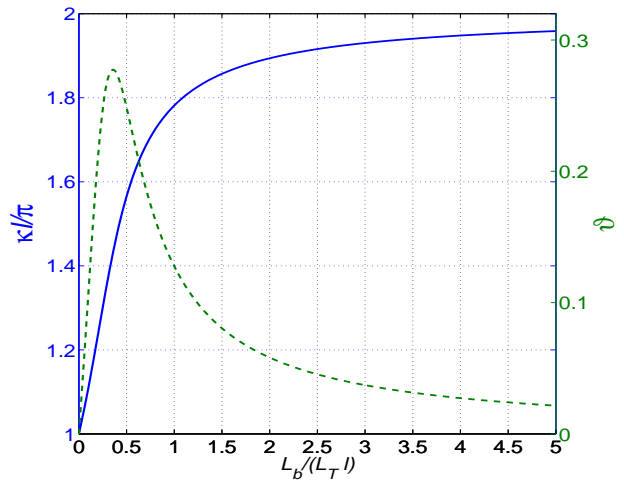}%
\caption{(color online) The factor $\kappa l$ (blue solid line) and
$\vartheta$ \ (green dashed line) calculated using Eq. (\ref{cot(kappa*l/2)})
as a function $L_{b}/L_{T}l$.}%
\label{vartheta}%
\end{center}
\end{figure}

Assuming that the only excited mode is the fundamental one, the capacitive
energy stored in the stripline resonator is found using Eqs. (\ref{V fm}) and
(\ref{cot(kappa*l/2)})%
\begin{equation}
\frac{C_{T}}{2}\int_{-l/2}^{l/2}\mathcal{V}^{2}\;\mathrm{d}y=\frac{C_{e}%
\dot{\varphi}^{2}}{2}\;,
\end{equation}
where $C_{e}$, which is given by%
\begin{equation}
C_{e}=\frac{C_{T}L_{T}}{2\kappa^{2}\vartheta L_{b}}\;, \label{C}%
\end{equation}
represents the effective capacitance of the stripline resonator. The factor
$\vartheta$, which is defined by%
\begin{equation}
\vartheta=-\frac{\sin\left(  \kappa l\right)  }{\kappa l+\sin\left(  \kappa
l\right)  }\ , \label{vartheta=}%
\end{equation}
can be calculated by solving numerically Eq. (\ref{cot(kappa*l/2)}) (see Fig.
\ref{vartheta}).%
Similarly, the inductive energy stored in the resonator, excluding the energy
stored in the lumped inductor $L_{b}$ at $y=0$, is found using Eqs.
(\ref{I fm}) and (\ref{cot(kappa*l/2)})%
\begin{equation}
\frac{L_{T}}{2}\int_{-l/2}^{l/2}\mathcal{I}^{2}\;\mathrm{d}y=\frac{\varphi
^{2}}{2L_{e}}\;, \label{ind energy ex}%
\end{equation}
where $L_{e}$, which is given by%
\begin{equation}
\frac{1}{L_{e}}=C_{e}\omega_{e}^{2}+\frac{1}{L_{b}}\;,
\end{equation}
represents the effective inductance of the stripline resonator excluding the
lumped element at $y=0$, and where%
\begin{equation}
\omega_{e}=\frac{\kappa}{\sqrt{L_{T}C_{T}}}\;. \label{omega_e}%
\end{equation}

The inductive energy stored in the RF SQUID loop and the lumped inductor
$L_{b}$ is calculated using Eqs. (\ref{I_s}) and (\ref{I_b})%
\begin{align}
&  \frac{1}{2}\left(
\begin{array}
[c]{cc}%
I_{s} & I_{b}%
\end{array}
\right)  \left(
\begin{array}
[c]{cc}%
\Lambda & M\\
M & L_{b}%
\end{array}
\right)  \left(
\begin{array}
[c]{c}%
I_{s}\\
I_{b}%
\end{array}
\right) \nonumber\\
&  =\frac{\varphi^{2}}{2L_{b}}+\frac{\left(  \Phi-\Phi_{e}-\frac{M\varphi
}{L_{b}}\right)  ^{2}}{2\Lambda\left(  1-\alpha_{M}^{2}\right)  }%
\;.\nonumber\\
&
\end{align}

\section{Lagrangian and Hamiltonian of the Closed System}

Here we derive a Lagrangian for the closed system consisting of the
nanomechanical resonator, stripline resonator and the RF SQUID. The effect of
damping will be later taken into account by introducing coupling to thermal
baths. The Lagrangian of the closed system is expressed as a function of $x$,
$\varphi$ and $\Phi$ and their time derivatives (denoted by overdot)%
\begin{equation}
\mathcal{L}=\frac{m\dot{x}^{2}}{2}+\frac{C_{e}\dot{\varphi}^{2}}{2}%
+\frac{C_{J}\dot{\Phi}^{2}}{2}-U_{0}-U_{1}\ ,
\end{equation}
where the potential terms are given by%
\begin{subequations}
\begin{align}
U_{0}  &  =\frac{m\omega_{m}^{2}x^{2}}{2}+\frac{C_{e}\omega_{e}^{2}\varphi
^{2}}{2}\ ,\\
U_{1}  &  =\frac{\left(  \Phi-\Phi_{e}-\frac{M\varphi}{L_{b}}\right)  ^{2}%
}{2\Lambda\left(  1-\alpha_{M}^{2}\right)  }-\frac{\Phi_{0}I_{c}\cos\frac
{2\pi\Phi}{\Phi_{0}}}{2\pi}\ .
\end{align}

Using Eqs. (\ref{Phi}), (\ref{I_s}), (\ref{I_b}) and (\ref{alpha_M}), the
corresponding Euler - Lagrange equations can be expressed as%
\end{subequations}
\begin{subequations}
\begin{align}
m\ddot{x}+m\omega_{m}^{2}x-I_{s}I_{b}\frac{\mathrm{d}M}{\mathrm{d}x}  &
=0\ ,\label{x dot dot}\\
C_{e}\ddot{\varphi}+\frac{\varphi}{L_{e}}+I_{b}  &
=0\ ,\label{varphi dot dot}\\
C_{J}\ddot{\Phi}+I_{s}+I_{c}\sin\frac{2\pi\Phi}{\Phi_{0}}  &  =0\ .
\label{Phi dot dot}%
\end{align}
The interpretation of these equations of motion is straightforward. Eq.
(\ref{x dot dot}) is Newton's 2nd law for the mechanical resonator, where the
force is composed of the restoring elastic force $-m\omega_{m}^{2}x$ and the
term due to the dependence of $M$ on $x$. Eq. (\ref{varphi dot dot}) relates
the current in the suspended beam $I_{b}$ with the currents in the effective
capacitor $C_{e}$ and the effective inductor $L_{e}$. Eq. (\ref{Phi dot dot})
states that the circulating current $I_{s}$ equals the sum of the current
$I_{c}\sin\theta$ through the JJ and the current $C_{J}\dot{V}_{J}$ through
the capacitor, where the voltage $V_{J}$ across the JJ is given by the second
Josephson equation $V_{J}=\left(  \Phi_{0}/2\pi\right)  \dot{\theta}$.

The variables canonically conjugate to $x$, $\varphi$ and $\Phi$ are given by
$p=m\dot{x}$, $q=C_{e}\dot{\varphi}$ and $Q=C_{J}\dot{\Phi}$ respectively. The
Hamiltonian is given by%
\end{subequations}
\begin{equation}
\mathcal{H}=p\dot{x}+q\dot{\varphi}+Q\dot{\Phi}-\mathcal{L}=\mathcal{H}%
_{0}+\mathcal{H}_{1}\ , \label{Hamiltonian}%
\end{equation}
where%
\begin{equation}
\mathcal{H}_{0}=\frac{p^{2}}{2m}+\frac{q^{2}}{2C_{e}}+U_{0}\ , \label{H_0}%
\end{equation}%
\begin{equation}
\mathcal{H}_{1}=\frac{Q^{2}}{2C_{J}}+U_{1}\ . \label{H_1}%
\end{equation}
Quantization is achieved by regarding the variables $\left\{  x,p,\varphi
,q,\Phi,Q\right\}  $ as Hermitian operators satisfying Bose commutation relations.

\section{Adiabatic Approximation}

As a basis for expanding the general solution we use the eigenvectors of the
following Schr\"{o}dinger equation%
\begin{equation}
\mathcal{H}_{1}\left\vert n\left(  x,\varphi\right)  \right\rangle
=\varepsilon_{n}\left(  x,\varphi\right)  \left\vert n\left(  x,\varphi
\right)  \right\rangle \ , \label{e.v H_1}%
\end{equation}
where $x$ and $\varphi$ are treated here as parameters (rather than degrees of
freedom). The local eigen-vectors are assumed to be orthonormal%
\begin{equation}
\left\langle m\left(  x,\varphi\right)  |n\left(  x,\varphi\right)
\right\rangle =\delta_{nm}\ .
\end{equation}

The eigenenergies $\varepsilon_{n}\left(  x,\varphi\right)  $ and the
associated wavefunctions $\varphi_{n}$ are found by solving the following
Schr\"{o}dinger equation%
\begin{equation}
\left(  -\beta_{C}\frac{\partial^{2}}{\partial\phi^{2}}+u\right)  \varphi
_{n}=\frac{\varepsilon_{n}}{E_{0}}\varphi_{n}\ . \label{Scrodinger phi}%
\end{equation}
where%
\begin{align}
u  &  =\frac{\left(  \phi-\phi_{0}\right)  ^{2}}{1-\alpha_{M}^{2}}+2\beta
_{L}\cos\phi\ ,\label{u}\\
\phi &  =\frac{2\pi\Phi}{\Phi_{0}}-\pi\ ,\label{phi=}\\
\phi_{0}  &  =\frac{2\pi\Phi_{e}}{\Phi_{0}}+\frac{2\pi M\left(  x\right)
\varphi}{\Phi_{0}L_{b}}-\pi\ ,\label{phi_0}\\
\beta_{L}  &  =\frac{2\pi\Lambda I_{c}}{\Phi_{0}}\ ,\\
\beta_{C}  &  =\frac{2e^{2}}{C_{J}E_{0}}\ ,\\
E_{0}  &  =\frac{\Phi_{0}^{2}}{8\pi^{2}\Lambda}\ .
\end{align}

The total wave function is expanded as%
\begin{equation}
\psi=\sum_{n}\xi_{n}\left(  x,\varphi,t\right)  \left\vert n\right\rangle \ .
\label{psi_adi}%
\end{equation}
In the adiabatic approximation \cite{Moody_160} the time evolution of the
coefficients $\xi_{n}$ is governed by the following set of decoupled equations
of motion%
\begin{equation}
\left[  \mathcal{H}_{0}+\varepsilon_{n}\left(  x,\varphi\right)  \right]
\xi_{n}=i\hbar\dot{\xi}_{n}\ .
\end{equation}
Note that in the present case the geometrical vector potential
\cite{Moody_160} vanishes since the wavefunctions $\varphi_{n}\left(
\phi\right)  $ can be chosen to be real. The validity of the adiabatic
approximation will be discussed below.

\section{Two level approximation}

In what follows we focus on the case where $\left\vert \phi_{0}\right\vert
\ll1$ and $\beta_{L}\left(  1-\alpha_{M}^{2}\right)  >1$. In this case the
adiabatic potential $u\left(  \phi\right)  $ given by Eq. (\ref{u}) contains
two wells separated by a barrier near $\phi=0$. At low temperatures only the
two lowest energy levels contribute. In this limit the local Hamiltonian
$\mathcal{H}_{1}$ can be expressed in the basis of the states $\left\vert
\curvearrowleft\right\rangle $ and $\left\vert \curvearrowright\right\rangle
$, representing localized states in the left and right well respectively
having opposite circulating currents. In this basis, $\mathcal{H}_{1}$ can be
expressed using Pauli's matrices%
\begin{equation}
\mathcal{H}_{1}\dot{=}\eta\phi_{0}\sigma_{z}+\Delta\sigma_{x}\ .
\label{2-level H_1}%
\end{equation}
The real parameters $\eta$ and $\Delta$ can be determined by solving
numerically the Schr\"{o}dinger equation (\ref{Scrodinger phi})
\cite{Buks_174504}. The eigenvectors and eigenenergies are denoted as
$\mathcal{H}_{1}\left\vert \pm\right\rangle =\varepsilon_{\pm}\left\vert
\pm\right\rangle $, where%
\begin{equation}
\varepsilon_{\pm}=\pm\sqrt{\eta^{2}\phi_{0}^{2}+\Delta^{2}}\ .
\label{epsilon_pm}%
\end{equation}

\section{Rotating Wave Approximation}

Consider the case where $\Phi_{e}=\Phi_{0}/2$, that is%
\begin{equation}
\phi_{0}=\frac{2\pi M\left(  x\right)  \varphi}{\Phi_{0}L_{b}}\;,
\end{equation}
and assume that adiabaticity holds and that the RF SQUID remains in its lowest
energy state. In this case, as can be seen from Eq. (\ref{epsilon_pm}),
expansion of $\varepsilon_{-}\left(  x,\varphi\right)  $ yields only even
powers of $\varphi$. These even powers of $\varphi$ can be expressed in terms
of the annihilation operator%
\begin{equation}
A_{e}=\frac{e^{i\omega_{e}t}}{\sqrt{2\hbar}}\left(  \sqrt{C_{e}\omega_{e}%
}\varphi+\frac{i}{\sqrt{C_{e}\omega_{e}}}q\right)  \
\end{equation}
and its Hermitian conjugate $A_{e}^{\dag}$, yielding terms oscillating at
frequencies $2n\omega_{e}$, where $n$ is integer. In the RWA such terms are
neglected unless $n=0$ since the effect of the oscillating terms on the
dynamics on a time scale much longer than a typical oscillation period is
negligibly small \cite{Santamore_144301}. Moreover, constant terms in the
Hamiltonian are disregarded since they only give rise to a global phase
factor. Displacement detection is performed by externally driving the
fundamental mode of the stripline resonator. To study the effect of
nonlinearity to lowest order we keep terms up to 4th order in $\varphi$. On
the other hand, since the mechanical displacement is assumed to be very small,
we keep terms up to 1st order only in $x$. Thus, in the RWA the Hamiltonian
$\mathcal{H}_{0}+\varepsilon_{-}$ is given by%
\begin{equation}
\mathcal{H}_{\mathrm{RWA}}=\hbar\omega_{m}N_{m}+\hbar\omega_{0}\left(
x\right)  N_{e}+\hbar KN_{e}^{2}\;,
\end{equation}
where $N_{m}$ and $N_{e}=A_{e}^{\dag}A_{e}$ are number operators of the
mechanical and stripline resonators respectively,%
\begin{align}
K  &  =\frac{3\Delta}{4\hbar}\left(  \frac{\eta}{\Delta}\frac{2\pi M_{0}}%
{\Phi_{0}L_{b}}\sqrt{\frac{\hbar}{2C_{e}\omega_{e}}}\right)  ^{4}%
\;,\label{K}\\
\omega_{0}\left(  x\right)   &  =\omega_{e}-\Omega_{2}\left(  1+2\frac
{\mathrm{d\log}M}{\mathrm{d}x}x\right)  \;,\label{Omega(x)}\\
\Omega_{2}  &  =\frac{\Delta}{\hbar}\left(  \frac{\eta}{\Delta}\frac{2\pi
M_{0}}{\Phi_{0}L_{b}}\sqrt{\frac{\hbar}{2C_{e}\omega_{e}}}\right)  ^{2}\;,
\label{Omega_2}%
\end{align}
and $M_{0}=M\left(  0\right)  $.

\section{Homodyne Detection}

The stripline resonator is weakly coupled (with a coupling constant
$\gamma_{1}$) to a semi infinite feedline, which guides the input and output
RF signals. To model the effect of dissipation (both linear and nonlinear), we
add two fictitious semi infinite transmission lines to the model, which allow
energy escape from the resonator. The first transmission line is linearly
coupled to the resonator with a coupling constant $\gamma_{2}$, and the second
one is nonlinearly coupled with a coupling constant $\gamma_{3}$
\cite{Yurke_5054}.

The dependence of $\omega_{0}$ on $x$ can be exploited for displacement
detection. This is achieved by exciting the fundamental mode of the stripline
resonator by launching into the feedline a monochromatic input pump signal
having a real amplitude $b_{in}$ and an angular frequency $\omega_{p}$ close
to the resonance frequency $\omega_{0}$. The output signal $c_{out}$ reflected
off the resonator is measured using homodyne detection, which is performed by
employing a balance mixing with an intense local oscillator having the same
frequency as the pump frequency $\omega_{p}$, and an adjustable phase
$\phi_{\mathrm{LO}}$. That is, the normalized (with respect to the amplitude
of the local oscillator) output signal of the homodyne detector is given by
\begin{equation}
X_{\phi_{\mathrm{LO}}}=c_{out}^{\dagger}e^{-i\phi_{\mathrm{LO}}}%
+c_{out}e^{i\phi_{\mathrm{LO}}}\ . \label{X_phi_LO}%
\end{equation}

To proceed, we employ below some results of Ref. \cite{Yurke_5054}, which has
studied a similar case of homodyne detection of a driven nonlinear resonator.

\subsection{Equation of Motion}

Using the standard method of Gardiner and Collett \cite{Gardiner_3761}, and
applying a transformation to a reference frame rotating at angular frequency
$\omega_{p}$%
\begin{equation}
A_{e}=Ce^{-i\omega_{p}t}\ ,
\end{equation}
yield the following equation for the operator $C$%
\begin{equation}
\frac{\mathrm{d}C}{\mathrm{d}t}+\Theta=F\left(  t\right)  \ , \label{dC/dt}%
\end{equation}
where%
\begin{align}
\Theta\left(  C,C^{\dagger}\right)   &  =\left[  \gamma+i\left(  \omega
_{0}-\omega_{p}\right)  +\left(  iK+\gamma_{3}\right)  C^{\dagger}C\right]
C\nonumber\\
&  +i\sqrt{2\gamma_{1}}b_{in}e^{i\phi_{1}}\ ,\nonumber\\
&  \label{Theta(C,C+)}%
\end{align}
$\gamma=\gamma_{1}+\gamma_{2}$, $\phi_{1}$ is the (real) phase shift of
transmission from the feedline into the resonator, and $F\left(  t\right)  $
is a noise term, having a vanishing average $\left\langle F\left(  t\right)
\right\rangle =0$, and an autocorrelation function, which is determined by
assuming that the three semi-infinite transmission lines are at thermal
equilibrium at temperatures $T_{1}$, $T_{2}$ and $T_{3}$ respectively.

\subsection{Linearization}

Let $C=B+c$, where $B$ is a complex number for which%
\begin{equation}
\Theta\left(  B,B^{\ast}\right)  =0\ , \label{Theta(B,B^*)}%
\end{equation}
namely, $B$ is a steady state solution of Eq. (\ref{dC/dt}) for the noiseless
case $F=0$. When the noise term $F$ can be considered as small, one can find
an equation of motion for the fluctuation around $B$ by linearizing Eq.
(\ref{dC/dt})%
\begin{equation}
\frac{\mathrm{d}c}{\mathrm{d}t}+Wc+Vc^{\dagger}=F\ , \label{dc/dt}%
\end{equation}
where%
\begin{equation}
W=\left.  \frac{\partial\Theta}{\partial C}\right\vert _{C=B}=\gamma+i\left(
\omega_{0}-\omega_{p}\right)  +2\left(  iK+\gamma_{3}\right)  B^{\ast}B\ ,
\label{W}%
\end{equation}
and%
\begin{equation}
V=\left.  \frac{\partial\Theta}{\partial C^{\dagger}}\right\vert
_{C=B}=\left(  iK+\gamma_{3}\right)  B^{2}\ .
\end{equation}

\subsection{Onset of Bistability Point}

In general, for any fixed value of the driving amplitude $b_{in}$, Eq.
(\ref{Theta(B,B^*)}) can be expressed as a relation between $\left\vert
B\right\vert ^{2}$ and $\omega_{p}$. When $b_{in}$ is sufficiently large the
response of the resonator becomes bistable, that is $\left\vert B\right\vert
^{2}$ becomes a multi-valued function of $\omega_{p}$. The onset of
bistability point is defined as the point for which%
\begin{align}
\frac{\partial\omega_{p}}{\partial\left\vert B\right\vert ^{2}}  &  =0\ ,\\
\frac{\partial^{2}\omega_{p}}{\partial\left(  \left\vert B\right\vert
^{2}\right)  ^{2}}  &  =0\ .
\end{align}
Such a point occurs only if the nonlinear damping is sufficiently small
\cite{Yurke_5054}, namely, only when the following condition holds%
\begin{equation}
\left\vert K\right\vert >\sqrt{3}\gamma_{3}\ .
\end{equation}
At the onset of bistability point the drive frequency and amplitude are given
by%
\begin{equation}
\left(  \omega_{p}-\omega_{0}\right)  _{c}=\gamma\frac{K}{\left\vert
K\right\vert }\left[  \frac{4\gamma_{3}|K|+\sqrt{3}\left(  K^{2}+\gamma
_{3}^{2}\right)  }{K^{2}-3\gamma_{3}^{2}}\right]  \ ,
\end{equation}%
\begin{equation}
\left(  b_{in}\right)  _{c}=\frac{8}{3\sqrt{3}}\frac{\gamma^{3}(K^{2}%
+\gamma_{3}^{2})}{\left(  \left\vert K\right\vert -\sqrt{3}\gamma_{3}\right)
^{3}}\ , \label{b_in,c}%
\end{equation}
and the resonator mode amplitude is%
\begin{equation}
\left\vert B\right\vert _{c}^{2}=\frac{2\gamma}{\sqrt{3}\left(  \left\vert
K\right\vert -\sqrt{3}\gamma_{3}\right)  }\ . \label{|B|_c}%
\end{equation}

\subsection{Ring-Down Time}

The solution of the equation of motion (\ref{dc/dt}) can be expressed as
\cite{Yurke_5054}%
\begin{equation}
c\left(  t\right)  =\int_{-\infty}^{\infty}\mathrm{d}t^{\prime}G\left(
t-t^{\prime}\right)  \Gamma\left(  t^{\prime}\right)  \ ,
\end{equation}
where%
\begin{equation}
\Gamma\left(  t\right)  =\frac{\mathrm{d}F\left(  t\right)  }{dt}+W^{\ast
}F\left(  t\right)  -VF^{\dagger}\left(  t\right)  \ .
\end{equation}
The propagator is given by%
\begin{equation}
G\left(  t\right)  =u\left(  t\right)  \frac{e^{-\lambda_{0}t}-e^{\lambda
_{1}t}}{\lambda_{1}-\lambda_{0}}\ ,
\end{equation}
where $u(t)$ is the unit step function, and the Lyapunov exponents
$\lambda_{0}$ and $\lambda_{1}$ are the eigenvalues of the homogeneous
equation, which satisfy%
\begin{equation}
\lambda_{0}+\lambda_{1}=2W^{\prime}\ , \label{lam_0+lam_1}%
\end{equation}%
\begin{equation}
\lambda_{0}\lambda_{1}=|W|^{2}-|V|^{2}\ , \label{lam_0*lam_1}%
\end{equation}
where $W^{\prime}$ is the real part of $W$. Thus one has%
\begin{equation}
\lambda_{0,1}=W^{\prime}\left(  1\pm\sqrt{1+\frac{|W|^{2}}{\left(  W^{\prime
}\right)  ^{2}}\left(  \zeta^{2}-1\right)  }\right)  \ ,
\end{equation}
where%
\begin{equation}
\zeta=\left\vert \frac{V}{W}\right\vert \ .
\end{equation}

We chose to characterize the ring-down time scale as%
\begin{equation}
t_{\mathrm{RD}}=\lambda_{0}^{-1}+\lambda_{1}^{-1}=\frac{2W^{\prime}%
}{\left\vert W\right\vert ^{2}\left(  1-\zeta^{2}\right)  }\ .
\label{t_RD nonlinear}%
\end{equation}
Note that in the limit $\zeta\rightarrow1$ slowing down occurs and
$t_{\mathrm{RD}}\rightarrow\infty$. This limit corresponds to the case of
operating the resonator near a jump point, close to the edge of the
bistability region. On the other hand, the limit $\zeta=0$ corresponds to the
linear case, for which $t_{\mathrm{RD}}$ at resonance ($\omega_{p}=\omega_{0}%
$) is given by Eq. (\ref{t_RD linear}).

\section{Displacement Sensitivity}

Consider a measurement in which $X_{\phi_{\mathrm{LO}}}\left(  t\right)  $ is
monitored in the time interval $\left[  0,\tau\right]  ,$ and the average
measured value is used to estimate the displacement $x$. Assuming the case
where $\tau$ is much longer than the characteristic correlation time of
$X_{\phi_{\mathrm{LO}}}\left(  t\right)  $, one finds that the minimum
detectable displacement is given by \cite{Ekinci_2682}%
\begin{equation}
\delta x=\left\vert \frac{\partial X_{0}}{\partial x}\right\vert ^{-1}\left(
\frac{2\pi}{\tau}\right)  ^{1/2}P_{X}^{1/2}\left(  0\right)  \ .
\end{equation}
Moreover, using Eq. (\ref{Omega(x)}), one finds the minimum detectable
displacement per square root of bandwidth, which is defined as $P_{x}=\left(
\tau/2\pi\right)  ^{1/2}\delta x$,$\ $is given by%
\begin{equation}
P_{x}=\left\vert 2\Omega_{2}\frac{\mathrm{d\log}M}{\mathrm{d}x}\right\vert
^{-1}R^{-1}P_{X}^{1/2}\left(  0\right)  \;,
\end{equation}
where $R$ is the responsivity with respect to a change in $\omega_{0}$, namely
$R=\left\vert \partial X_{0}/\partial\omega_{0}\right\vert $. To evaluate
$P_{x}$ we calculate below the responsivity $R$ and the spectral density
$P_{X}^{1/2}\left(  0\right)  $ using results, which were obtained in Ref.
\cite{Yurke_5054}.

\subsection{Responsivity}

In steady state, when input noise is disregarded, the amplitude of the
fundamental mode of the stripline resonator is expressed as $Be^{-i\omega
_{p}t}$, where the complex number $B$ is found by solving Eq.
(\ref{Theta(B,B^*)}), and the average output signal is expressed as
$b_{out}e^{-i\omega_{p}t}$ ($b_{out}$ is in general complex). The amplitude of
the output signal in the feedline $b_{out}$ is related to the input signal
$b_{in}$ and the mode amplitude $B$ by an input-output relation
\cite{Gardiner_3761}%
\begin{equation}
b_{out}=b_{in}-i\sqrt{2\gamma_{1}}e^{-i\phi_{1}}B\ . \label{b_out}%
\end{equation}

Consider a small change in the resonance frequency $\delta\omega_{0}$. The
resultant change in the steady state mode amplitude $\delta B$ can be
calculated using Eq. (\ref{Theta(B,B^*)})%
\begin{equation}
i\left(  \delta\omega_{0}\right)  B+W\delta B+V\left(  \delta B\right)
^{\ast}=0\ . \label{eq for delta B}%
\end{equation}
The solution of this equation together with Eqs. (\ref{X_phi_LO}) and
(\ref{b_out}) allows calculating the responsivity%
\begin{align}
R  &  =\frac{2\sqrt{2\gamma_{1}}\left\vert B\right\vert }{\left\vert
W\right\vert \left(  1-\zeta^{2}\right)  }\left\vert \sin\left(  \phi_{t}%
+\phi_{C}\right)  +\zeta\sin\left(  \phi_{t}-\phi_{C}\right)  \right\vert
\ ,\nonumber\\
&
\end{align}
where%
\begin{align}
\phi_{t}  &  =\frac{2\phi_{\mathrm{LO}}-\phi_{W}+\phi_{V}-2\phi_{1}}{2}\;,\\
\phi_{C}  &  =\frac{2\phi_{B}-\phi_{W}-\phi_{V}-\pi}{2}\;,
\end{align}
$e^{i\phi_{B}}=B/\left\vert B\right\vert $, $e^{i\phi_{W}}=W/\left\vert
W\right\vert $ and $e^{i\phi_{V}}=V/\left\vert V\right\vert $.

\subsection{Spectral Density}

The zero frequency spectral density $P_{X}\left(  0\right)  $ was calculated
in Ref. \cite{Yurke_5054}%
\begin{align}
P_{X}\left(  0\right)   &  =\left\vert \frac{2\gamma_{1}\left(  1-\zeta
e^{2i\phi_{t}}\right)  }{\left\vert W\right\vert \left(  1-\zeta^{2}\right)
}-e^{-i\phi_{W}}\right\vert ^{2}\coth\frac{\hbar\omega_{0}}{2k_{B}T_{1}%
}\nonumber\\
&  +\frac{\gamma_{2}}{\gamma_{1}}\left\vert \frac{2\gamma_{1}\left(  1-\zeta
e^{2i\phi_{t}}\right)  }{\left\vert W\right\vert \left(  1-\zeta^{2}\right)
}\right\vert ^{2}\coth\frac{\hbar\omega_{0}}{2k_{B}T_{2}}\nonumber\\
&  +\frac{2\gamma_{3}\left\vert B\right\vert ^{2}}{\gamma_{1}}\left\vert
\frac{2\gamma_{1}\left(  1-\zeta e^{2i\phi_{t}}\right)  }{\left\vert
W\right\vert \left(  1-\zeta^{2}\right)  }\right\vert ^{2}\coth\frac
{\hbar\omega_{0}}{2k_{B}T_{3}}\ .\nonumber\\
&  \label{P_phi_LO_(0)}%
\end{align}

\subsection{The Linear Case}

In this case $K=\gamma_{3}=0$, $W=i\left(  \omega_{0}-\omega_{p}\right)
+\gamma$ and $V=0$, and consequently $\zeta=0$. Thus, the responsivity is
given by%
\begin{equation}
R=\frac{2\sqrt{2\gamma_{1}}\left\vert B\right\vert \left\vert \sin\left(
\phi_{t}+\phi_{C}\right)  \right\vert }{\left\vert W\right\vert }\ .
\end{equation}
Moreover, for the case where $T_{1}=T_{2}$, Eq. (\ref{P_phi_LO_(0)}) becomes%
\begin{equation}
P_{X}\left(  0\right)  =\coth\frac{\hbar\omega_{0}}{2k_{B}T_{1}}\ .
\end{equation}
The largest responsivity is obtained at resonance, namely when $\omega
_{p}=\omega_{0}$, and when the homodyne detector measures the phase of
oscillations, namely, when $\left\vert \sin\left(  \phi_{t}+\phi_{C}\right)
\right\vert =1$. For this case $P_{x}$ obtains its smallest value, which is
denoted as $P_{x0}$, and is given by%
\begin{equation}
P_{x0}=\left\vert 2\Omega_{2}\frac{\mathrm{d\log}M}{\mathrm{d}x}\right\vert
^{-1}\left(  \frac{\gamma^{2}}{8\gamma_{1}\left\vert B\right\vert ^{2}}%
\coth\frac{\hbar\omega_{0}}{2k_{B}T_{1}}\right)  ^{1/2}\;.
\end{equation}
Note that this result coincides with the SLL value $P_{x}^{\mathrm{SLL}}$
given by Eq. (\ref{P_x linear}) provided that the quality factor in Eq.
(\ref{P_x linear}) is taken to be given by $Q=2\omega_{0}\gamma_{1}/\gamma
^{2}$. That is, $P_{x0}\simeq P_{x}^{\mathrm{SLL}}$ in the limit of strongly
overcoupled resonator, namely when the damping of the resonator is dominated
by the coupling to the feedline ($\gamma_{1}\simeq\gamma$).

The response of the stripline resonator is approximately linear only when
$\left\vert B\right\vert $ is much smaller than the critical value
corresponding to the onset of nonlinear bistability. Using this critical value
$\left\vert B\right\vert _{c}$, which is given by Eq. (\ref{|B|_c}), and using
Eqs. (\ref{K}) and (\ref{Omega_2}), one finds that the smallest possible value
of $P_{x0}$ in this regime is roughly given by%
\begin{equation}
P_{x0,c}\simeq\frac{0.14\left\vert \frac{\mathrm{d\log}M}{\mathrm{d}%
x}\right\vert ^{-1}}{\sqrt{\Delta/\hbar}}\sqrt{\frac{\gamma}{\gamma_{1}%
}\left(  1-\frac{\sqrt{3}\gamma_{3}}{\left\vert K\right\vert }\right)
\coth\frac{\hbar\omega_{0}}{2k_{B}T_{1}}}\;. \label{P_xo,c}%
\end{equation}

\subsection{The General Case}

The minimum detectable displacement per square root of bandwidth in the
general case can be written as
\begin{equation}
P_{x}=P_{x}^{\mathrm{SLL}}g\left(  \phi_{t}\right)  \;,
\end{equation}
where%
\begin{equation}
g\left(  \phi_{t}\right)  =\frac{\frac{\left\vert W\right\vert }{\gamma
}\left[  \frac{\left(  1-\zeta^{2}\right)  ^{2}P_{X}\left(  0\right)  }%
{\coth\frac{\hbar\omega_{0}}{2k_{B}T_{1}}}\right]  ^{1/2}}{\left\vert
\sin\left(  \phi_{t}+\phi_{C}\right)  +\zeta\sin\left(  \phi_{t}-\phi
_{C}\right)  \right\vert }\;, \label{g(phi_t)}%
\end{equation}
and $P_{X}\left(  0\right)  $ is given by Eq. (\ref{P_phi_LO_(0)}). The
function $g\left(  \phi_{t}\right)  $ is periodic with a period $\pi$. Its
minimum value is denoted as $g_{\min}$. Beating the SLL given by Eq.
(\ref{P_x linear}) is achieved when $g_{\min}$ is made smaller than unity.

\section{Validity of Approximations}

In this section we examine the conditions, which are required to justify the
approximations made, and determine the range of validity of our results.
Clearly, our analysis breaks down if the driving amplitude $b_{in}$ is made
sufficiently large. In this case both the adiabatic approximation and the
assumption that back-reaction effects are negligibly small will become
invalid. The range of validity of the adiabatic approximation is examined
below by estimating the rate of Zener transitions. Moreover, we study below
the conditions under which back-reaction effects acting back on the mechanical
resonator play an important role. Using these results we derive conditions for
the validity of the above mentioned approximations. These conditions are then
examined for the case where the system is driven to the onset of nonlinear
bistability. This analysis allows us to determine whether the device can be
operated in the regime of nonlinear bistability, where the effects of
bifurcation amplification
\cite{Wiesenfeld_629,Dykman_1198,Krommer_101,Savel'ev_056136,Chan_0603037,Almog_213509}
and noise squeezing can be exploited
\cite{Yurke_5054,Rugar_699,Almog_0607055,Segev_0607262}, without, however,
violating the adiabatic approximation and without inducing strong
back-reaction effects.

\subsection{Adiabatic Condition}

As before, consider the case where the externally applied flux is given by
$\Phi_{e}=\Phi_{0}/2$, and the stripline resonator is driven close to the
onset of nonlinear bistability, where the number of photons approaches the
critical value given by Eq. (\ref{|B|_c}), and assume for simplicity the case
where $\gamma_{3}\ll\left\vert K\right\vert $. Using Refs.
\cite{Buks_174504,Buks_628} one finds that adiabaticity holds, namely Zener
transitions between adiabatic states are unlikely, provided that%
\begin{equation}
\frac{\pi\Delta^{2}}{\eta\hbar\Gamma_{c}}\gtrsim1\ , \label{adi_cond}%
\end{equation}
where%
\begin{equation}
\Gamma_{c}=\frac{2\pi M_{0}}{\Phi_{0}L_{b}}\sqrt{\frac{2\hbar\omega
_{e}\left\vert B\right\vert _{c}^{2}}{C_{e}}}\ .
\end{equation}

\subsection{Back Reaction}

Back reaction of the driven stripline resonator results in a force noise
acting on the mechanical resonator and a renormalization of the mechanical
resonance frequency $\omega_{m}$ and the damping rate $\gamma_{m}$
\cite{Braginsky_2002,Dykman_1306}. The renormalized values, denoted as
$\omega_{m}^{\mathrm{eff}}$ and $\gamma_{m}^{\mathrm{eff}}$, are expressed
using the renormalization factors $R_{\mathrm{f}}$ and $R_{\mathrm{d}}$%
\begin{subequations}
\begin{align}
R_{\mathrm{f}} &  =\frac{\omega_{m}^{\mathrm{eff}}-\omega_{m}}{\gamma_{m}%
}\;,\\
R_{\mathrm{d}} &  =\frac{\gamma_{m}^{\mathrm{eff}}-\gamma_{m}}{\gamma_{m}}\;,
\end{align}
which where calculated in Ref. \cite{Blencowe_0704_0457}%
\end{subequations}
\begin{subequations}
\begin{align}
R_{\mathrm{f}} &  =\chi_{\mathrm{f}}Q_{m}R_{0}\;,\\
R_{\mathrm{d}} &  =\chi_{\mathrm{d}}Q_{m}R_{0}\;,
\end{align}
where $Q_{m}=\omega_{m}/\gamma_{m}$,%
\end{subequations}
\begin{equation}
R_{0}=\frac{8\hbar\Omega_{2}^{2}\left\vert B\right\vert ^{2}\left(
\frac{\mathrm{d\log}M}{\mathrm{d}x}\right)  ^{2}}{m\omega_{m}^{3}%
}\;,\label{R_0}%
\end{equation}%
\begin{subequations}
\begin{align}
\chi_{\mathrm{f}} &  =\frac{\frac{1}{2}\frac{\Delta\omega}{\omega_{m}}\left[
\left(  \frac{\gamma}{\omega_{m}}\right)  ^{2}-1+\left(  \frac{\Delta\omega
}{\omega_{m}}\right)  ^{2}\right]  }{\left[  \left(  \frac{\gamma}{\omega_{m}%
}\right)  ^{2}+\left(  1+\frac{\Delta\omega}{\omega_{m}}\right)  ^{2}\right]
\left[  \left(  \frac{\gamma}{\omega_{m}}\right)  ^{2}+\left(  1-\frac
{\Delta\omega}{\omega_{m}}\right)  ^{2}\right]  }\;,\label{chi_f}\\
\chi_{\mathrm{d}} &  =-\frac{\frac{\gamma}{\omega_{m}}\frac{\Delta\omega
}{\omega_{m}}}{\left[  \left(  \frac{\gamma}{\omega_{m}}\right)  ^{2}+\left(
1+\frac{\Delta\omega}{\omega_{m}}\right)  ^{2}\right]  \left[  \left(
\frac{\gamma}{\omega_{m}}\right)  ^{2}+\left(  1-\frac{\Delta\omega}%
{\omega_{m}}\right)  ^{2}\right]  }\;,\label{chi_d}%
\end{align}
and%
\end{subequations}
\begin{equation}
\Delta\omega=\omega_{e}-\Omega_{2}-\omega_{p}\;.
\end{equation}

We now wish to examine whether backreaction effects are important when the
stripline resonator is driven to the onset of nonlinear bistability. For
simplicity we neglect nonlinear damping, that is we take $\gamma_{3}=0$. Using
Eqs. (\ref{K}), (\ref{Omega_2}) and (\ref{|B|_c}), one finds that the value of
$R_{0}$ [Eq. (\ref{R_0})] at the onset of nonlinear bistability, which is
denoted as $R_{0c}$, is given by%
\begin{equation}
R_{0c}=\frac{\gamma}{\omega_{m}}\frac{64\Delta\left(  \frac{\mathrm{d\log}%
M}{\mathrm{d}x}\right)  ^{2}}{3\sqrt{3}m\omega_{m}^{2}}\;. \label{R_0c}%
\end{equation}
It is straightforward to show that $\left\vert \chi_{\mathrm{f}}\right\vert
<\omega_{m}/\gamma$ and $\left\vert \chi_{\mathrm{d}}\right\vert <\omega
_{m}/\gamma$ for any value of $\Delta\omega$. Thus, back-reaction can be
considered as negligible when%
\begin{equation}
\frac{\omega_{m}^{2}}{\gamma\gamma_{m}}R_{0c}\ll1\;.
\label{back-reaction negligible}%
\end{equation}

\section{Example}

We examine below an example of a device having the following parameters:
$Z_{T}=50%
\operatorname{\Omega }%
$, $L_{b}/L_{T}l=1$, $\omega_{e}/2\pi=5%
\operatorname{GHz}%
$, $\omega_{e}/\gamma=10^{4}$, $\gamma_{2}=0.1\gamma_{1}$, and $\gamma
_{3}=0.1K/\sqrt{3}$ for the stripline resonator, $m=10^{-16}%
\operatorname{kg}%
$, $\omega_{m}/2\pi=0.01%
\operatorname{GHz}%
$, and $\omega_{m}/\gamma_{m}=10^{4}$ for the nanomechanical resonator,
$\Lambda=9.1\times10^{-10}%
\operatorname{H}%
$, $C_{J}=0.86\times10^{-14}%
\operatorname{F}%
$, and $I_{c}=0.44%
\operatorname{\mu A}%
$ for the RF SQUID, coupling terms $M_{0}/L_{b}=0.005$, $L_{b}/\Lambda=0.25$,
$\left\vert \mathrm{d\log}M/\mathrm{d}x\right\vert ^{-1}=100%
\operatorname{nm}%
$, and temperature $T=0.02%
\operatorname{K}%
$. These parameters for the stripline resonator \cite{Segev_1943}, for the
nanomechanical resonator \cite{Roukes_0008187}, and for the RF SQUID
\cite{Friedman_43,Koch_1216}, are within reach with present day technology.

Using FastHenry simulation program (www.fastfieldsolvers.com) we find that the
chosen value of $\Lambda$ corresponds, for example, to a loop made of Al
(penetration depth $50%
\operatorname{nm}%
$), having a square shape with edge length of $190%
\operatorname{\mu m}%
$, line width of $1%
\operatorname{\mu m}%
$ and film thickness of $50%
\operatorname{nm}%
$. The values of $C_{J}$ and $I_{c}$ correspond to a junction having a plasma
frequency of about $25%
\operatorname{GHz}%
$.

Using these values one finds $\beta_{L}=1.2$, $\beta_{C}=0.1$, and
$E_{0}/\hbar=560%
\operatorname{GHz}%
$. The values of $\beta_{L}$, $\beta_{C}$ and $\alpha_{M}$ are employed for
calculating numerically the eigenstates of Eq. (\ref{Scrodinger phi})
\cite{Buks_174504}. From these results one finds the parameters in the
two-level approximation of Hamiltonian $\mathcal{H}_{1}$ [Eq.
(\ref{2-level H_1})] $\eta=1.2E_{0}$ and $\Delta=0.073E_{0}$, and the
nonlinear terms $\Omega_{2}/2\pi=1.3%
\operatorname{MHz}%
$ and $K/2\pi=0.19%
\operatorname{kHz}%
$.

The validity of the adiabatic approximation is confirmed by evaluating the
factor in inequality (\ref{adi_cond})%
\begin{equation}
\frac{\pi\Delta^{2}}{\eta\hbar\Gamma_{c}}=12\ ,
\end{equation}
whereas, to confirm that back-reaction effects can be considered as negligible
we evaluate the factor in inequality (\ref{back-reaction negligible})%
\begin{equation}
\frac{\omega_{m}^{2}}{\gamma\gamma_{m}}R_{0c}=1.4\times10^{-4}\;.
\label{Q_m R_0c}%
\end{equation}

\begin{figure}
[ptb]
\begin{center}
\includegraphics[
height=4.6899in,
width=3.2396in
]%
{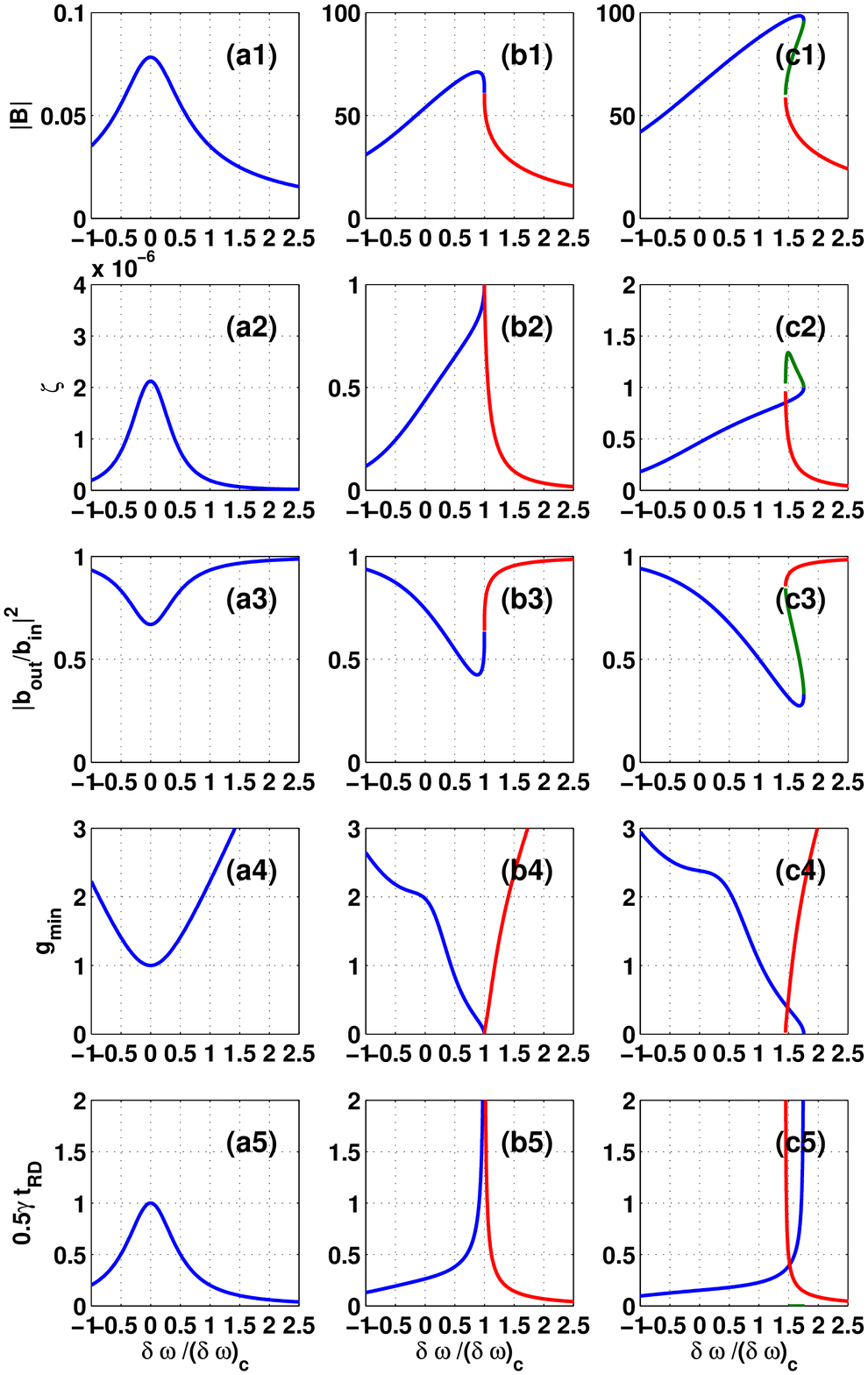}%
\caption{(Color online) Dependence of the mode amplitude $\left\vert
B\right\vert $, the factor $\zeta$, the reflection coefficient $\left\vert
b_{out}/b_{in}\right\vert ^{2}$, the enhancement factor $g_{\min}$, and the
ring-down time $t_{\mathrm{RD}}$ on the pump frequency. Panel (a) represents
the nearly linear case for which $b_{in}=0.001\left(  b_{in}\right)  _{c}$,
panel (b) the critical case for which $b_{in}=\left(  b_{in}\right)  _{c}$,
and panel (c) the over-critical case for which $b_{in}=1.5\left(
b_{in}\right)  _{c}$.}%
\label{Duffing}%
\end{center}
\end{figure}

Using Eq. (\ref{P_xo,c}) we calculate the smallest possible value of $P_{x0}$%
\begin{equation}
P_{x0,c}=6.9\times10^{-14}\frac{%
\operatorname{m}%
}{\sqrt{%
\operatorname{Hz}%
}}\;.
\end{equation}
A significant sensitivity enhancement, however, can be achieved by exploiting
nonlinearity. Figure (\ref{Duffing}) shows the dependence of the mode
amplitude $\left\vert B\right\vert $, the factor $\zeta$, the reflection
coefficient $\left\vert b_{out}/b_{in}\right\vert ^{2}$, the enhancement
factor $g_{\min}$, and the ring-down time $t_{\mathrm{RD}}$ on the pump
frequency. Panel (a) represents the nearly linear case for which
$b_{in}=0.001\left(  b_{in}\right)  _{c}$, panel (b) the critical case for
which $b_{in}=\left(  b_{in}\right)  _{c}$, and panel (c) the over-critical
case for which $b_{in}=1.5\left(  b_{in}\right)  _{c}$. As can be clearly seen
from Fig. (\ref{Duffing}) a significant sensitivity enhancement $g_{\min}\ll1$
can be achieved when driving the resonator close to a jump point. However, in
the same region, slowing down occurs, resulting in a long ring-down time
$t_{\mathrm{RD}}\gg2/\gamma$.%

\section{Discussion and Conclusions}

In the present paper we consider a resonant detection configuration, in which
the response of the resonator is measured by monitoring an output signal in
the feedline, which is reflected off the resonator. We show that operating in
the regime of nonlinear response may allow a significant enhancement in the
sensitivity by driving the stripline resonator near a jump point at the edge
of the region of bistability. The factor $g_{\min}$, which represents this
enhancement, can be made significantly smaller than unity in this limit.
However, this result is not general to all resonance detectors. Consider an
alternative configuration, in which, instead of measuring an output signal
reflected off the resonator, the response is measured by directly homodyning
the mode amplitude in the resonator. In the latter case, a similar analysis
yields that the enhancement factor $g_{\min}$ cannot be made smaller than
$0.5$ \cite{Buks_0606081}. Thus, to take full advantage of nonlinearity, it is
advisable to monitor the response of the resonator by measuring a reflected
off signal, rather than measuring an internal cavity signal.

As was discussed above, the largest sensitivity enhancement is obtained close
to the edge of the region of bistability, where $1-\zeta\ll1$. On the other
hand, in the very same region, the perturbative approach, which we employ to
study the response of the stripline resonator, becomes invalid since the
fluctuation around steady state, which is assumed to be small, is strongly
enhanced due to bifurcation amplification of input noise \cite{Yurke_5054}.
The integrated spectral density of the fluctuation, which was calculated in
Eq. (80) of Ref. \cite{Buks_0606081} (for the case $T_{1}=T_{2}=T_{3}$), can
be employed to determine the range of validity of the perturbative approach%
\begin{equation}
\frac{1}{1-\zeta}\coth\frac{\hbar\omega_{0}}{2k_{B}T_{1}}\ll\left\vert
B\right\vert ^{2}\;. \label{zeta inequality}%
\end{equation}
By applying this condition to the case of operating at the onset of
bistability point [see panel (b) of Fig. (\ref{Duffing})], one finds that the
smallest value of $g_{\min}$ in the region where inequality
(\ref{zeta inequality}) holds is $\simeq10^{-3}$ for the set of parameters
chosen in the above considered example . Thus a significant sensitivity
enhancement is achieved even when the region close to the jump, where the
perturbative analysis breaks down, is excluded. On the other hand, the
perturbative analysis cannot answer the question what is the smallest possible
value of $g_{\min}$. Further work is required in order to answer this question
and to properly describe the system very close to the edge of the region of
bistability, where nonlinear terms of higher orders become important.

In the present paper only the case where back-reaction effects do not play an
important role is considered. Thus our results are inapplicable when the
displacement sensitivity approaches the value corresponding to zero point
motion of the mechanical resonator $\sqrt{\hbar/m\omega_{m}\gamma_{m}}$
($1.6\times10^{-15}%
\operatorname{m}%
/\sqrt{%
\operatorname{Hz}%
}$ for the above considered example), and consequently quantum back-reaction
becomes important \cite{Caves_1817}. Moreover, the results are valid only when
inequality (\ref{back-reaction negligible}) is satisfied. To increase the
range of applicability of the theory and to study back-reaction effects, a
more general approach will be considered in a forthcoming paper
\cite{Nation_InProgress}.


\section*{Acknowledgment}

We are grateful to Bernard Yurke for reviewing the manuscript. This work is
partly supported by the US - Israel binational science foundation, Israel
science foundation, Devorah foundation, Poznanski foundation, Russel Berrie
nanotechnology institute, and by the Israeli ministry of science.

\newpage
\bibliographystyle{apsrev}
\bibliography{acompat,Eyal_Bib}

\end{document}